\documentclass[aps,prl,twocolumn,superscriptaddress,showpacs,reprint]{revtex4-1}

\usepackage{graphicx}

\newcommand{\be}{\begin{equation}}
\newcommand{\ee}{\end{equation}}
\newcommand{\ba}{\begin{eqnarray}}
\newcommand{\ea}{\end{eqnarray}}

\begin{document} 
%%%%%%%%%%%%%%%%%%%%%%%%%%%%%%%%%%%%%%%%%%%%%%%%%%%%%%%%%%%
%%%%%%%%%%%%%%%%%%%%%%%%%%%%%%%%%%%%%%%%%%%%%%%%%%%%%%%%%%%

\title{Rapid Cooling of the Neutron Star in Cassiopeia A \\
	Triggered by Neutron Superfluidity in Dense Matter} 
\author{\surname{Dany} Page}
\affiliation{Instituto de Astronom\'{\i}a, 
             Universidad Nacional Aut\'onoma de M\'exico, 
             Mexico D.F. 04510, Mexico}
\author{\surname{Madappa} Prakash}
\affiliation{Department of Physics and Astrononmy, 
                Ohio University,
                Athens, OH 45701-2979, USA}
\author{\surname{James M.} Lattimer}
\affiliation{Department of Physics and Astronomy, 
                State University of New York at Stony Brook,
                Stony Brook, NY-11794-3800, USA}
\author{\surname{Andrew W.} Steiner}
\affiliation{Joint Institute for Nuclear Astrophysics, 
        National Superconducting Cyclotron Laboratory and, \\ 
        Department of Physics and Astrononmy, 
        Michigan State University, East Lansing, MI 48824, USA}

%%%%%%%%%%%%%%%%%%%%%%%%%%%%%%%%%%%%%%%%%%%%%%%%%%%%%%%%%%%
\begin{abstract} 

We propose that the observed cooling of the neutron star in Cassiopeia A 
is due to enhanced neutrino emission from the recent onset of the breaking and formation
of  neutron Cooper pairs in the $^3$P$_2$ channel.  
We find that the critical temperature for this superfluid transition is $\simeq 0.5\times 10^9$ K.
The observed rapidity of the cooling implies that protons were 
already in a superconducting state with a larger critical temperature.
Our prediction that this cooling will continue for several decades at 
the present rate can be tested by continuous monitoring of this neutron star.

\end{abstract} 
%%%%%%%%%%%%%%%%%%%%%%%%%%%%%%%%%%%%%%%%%%%%%%%%%%%%%%%%%%%
\pacs{97.60.Jd, 95.30.Cq,, 26.60.-c} 

% 97.60.Jd   Neutron stars
% 95.30.Cq   Elementary particle processes
% 26.60.-c   Nuclear matter aspects of neutron stars 

\maketitle 

%%%%%%%%%%%%%%%%%%%%%%%%%%%%%%%%%%%%%%%%%%%%%%%%%%%%%%%%%%%
%%%%%%%%%%%%%%%%%%%%%%%%%%%%%%%%%%%%%%%%%%%%%%%%%%%%%%%%%%%

The neutron star in Cassiopeia~A (``Cas~A" hereafter),
discovered in 1999 in the {\em Chandra} first light observation
\cite{Tananbaum:1999kx} targeting the supernova remnant, 
is the youngest known in the Milky Way.  An association
with the historical supernova SN~1680 \cite{Ashworth:1980vn} gives
Cas~A an age of 330~yrs, in agreement with its kinematic age
\cite{Fesen:2006ys}. The distance to the remnant is estimated to
be $3.4^{+0.3}_{-0.1}$~kpc \cite{Reed:1995zr}.  
The thermal soft X-ray spectrum of
Cas~A is well fit by a non-magnetized carbon atmosphere model,
with a surface temperature of $2\times 10^6$~K
and an emitting radius of $8 - 17$~km \cite{Ho:2009fk}.
These results raise Cas A to the rank of the very few isolated neutron stars
with a well determined age and a reliable surface temperature,
thus allowing for detailed modeling of its thermal evolution and 
the determination of its interior properties \cite{Yakovlev:2010fk}. 

Analyzing archival data from 2000 to 2009,
Heinke \& Ho \cite{Heinke:2010xy} recently reported that Cas~A's surface
temperature has rapidly decreased from $2.12 \times 10^6$ to $2.04 \times 10^6$~K 
\cite{Note0,shternin2010}.
This rate of cooling is significantly larger than expected from the 
modified Urca (``MU'') process~\cite{Yakovlev:2004kx,Page:2006ly}
or a medium modified Urca \cite{grigorian2005}.
It is also unlikely to be due to any of the fast neutrino ($\nu$) emission processes 
(such as direct Urca processes from nucleons or hyperons, 
or $\nu-$emission from Bose condensates or gapless quark matter) 
since the visible effects of those become apparent over the 
thermal relaxation timescale of the crust ~\cite{Lattimer:1994fk}, i.e., $30-100$~yrs, 
much earlier than the age of Cas~A, and exhibit a slow evolution at later times.
We interpret Cas~A's cooling within the ``Minimal Cooling" paradigm \cite{Page:2004zr} 
and suggest it is due to the recent
triggering of enhanced neutrino emission resulting from the neutron
$^3P_2$ pairing in the star's core.
Our  numerical calculations and analytical analysis imply a critical
temperature $T_C \simeq 0.5\times 10^9$~K 
for the triplet neutron superfluidity.

The essence of the minimal cooling paradigm is the {\em a priori}
exclusion of all fast $\nu$-emission mechanisms, thus
restricting $\nu$-emission to the ``standard" MU process and 
nucleon bremsstrahlung processes~\cite{Page:2006ly}.  
However, effects of pairing, i.e., neutron
superfluidity and/or proton superconductivity, are included. At
temperatures just below the critical temperature $T_c$ of a pairing phase
transition, the continuous breaking and formation of Cooper pairs
\cite{Flowers:1976vn}, referred to as the ``PBF" process,
results in an enhanced neutrino emission.  
Calculations of $T_c$ for neutrons, $T_{cn}$, in
the $^3P_2$ channel relevant for neutron star cores,
are uncertain due to unsettled interactions \cite{Baldo98}
and medium effects which can either strongly suppress or increase the pairing
\cite{Schwenk:2004kx}.
Consequently, predictions range from vanishingly small to almost $10^{11}$ K
\cite{Page:2004zr}.  
The pairing gap is density ($\rho$) dependent, and the
resulting $T_{cn}(\rho)$ commonly exhibits a bell-shaped density profile.
Assuming the neutron star has an isothermal core at temperature $T$, the phase
transition will start when $T$ reaches, at some location in the star,
the maximum value of $T_{cn}(\rho)$: $T_C \equiv \max T_{cn}(\rho)$.
At that stage, neutrons in a thick shell go through the phase
transition and as $T$ decreases, this shell splits into two shells
which slowly drift toward the lower and higher density regions away
from the maximum of the bell-shaped profile.  If the neutron $^3P_2$
gap has the appropriate size, $\nu$-emission from the PBF process is
an order of magnitude more efficient than the MU process (see Fig.~20
of \cite{Page:2004zr} or Fig.~2 of \cite{Gusakov:2004ys}).

%++++++++++++++++++++++++++++++++++++++++++++++++++++++++++++++++++++++
\begin{figure}[t]
   \begin{center}
   \includegraphics[width=0.48\textwidth]{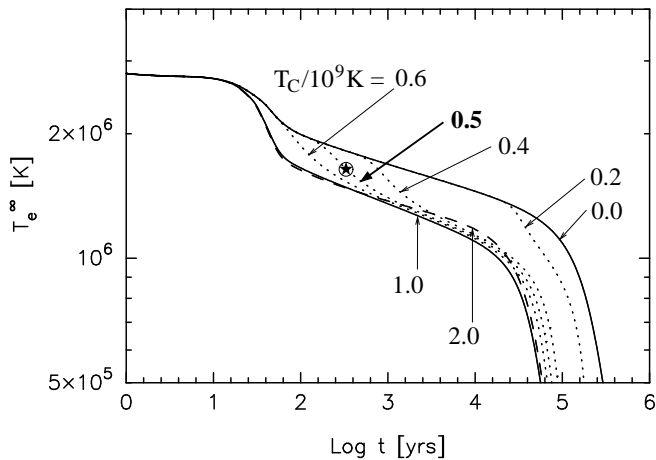}
  \end{center}
   \caption{Red-shifted effective temperature $T_e^\infty$ vs. time,
                  of an isolated neutron star for various
                  values of the maximum neutron $^3P_2$ pairing
                  critical temperature, $T_C$ (in units of $10^9$~K).  
                  The circled star shows the value for  Cas~A. 
	        Results shown are from
                  numerical solutions of the general relativistic
                  energy balance and heat transport equations
                  for a $1.4~{\rm M}_\odot$ star built from the APR
                  equation of state \cite{Akmal:1998qf}.
                  Refs.~\cite{Page:2004zr,Page:2009ly} describe the
                  microphysical inputs used.
                  These results are similar to those in Fig.~6 of
                  \cite{Page:2009ly}, but with the difference that the
                  upper layers of the envelope contain light elements
                  of total mass $\Delta M_\mathrm{light} = 5 \times
                  10^{-11} {\rm M}_\odot$, 
	       compatible with the
                  observation of carbon at the star's surface
                  \cite{Ho:2009fk,Heinke:2010xy}.  The rapid cooling
                  at ages $\sim$ 30-100 yrs is due to the thermal
                  relaxation of the crust. 
	       Thereafter, the stellar interior is isothermal.
                  }
   \label{Fig1} 
\end{figure}
%++++++++++++++++++++++++++++++++++++++++++++++++++++++++++++++++++++++

Implications of the size of the neutron $^3P_2$ pairing gap were considered 
in \cite{Page:2009ly} with the result that,  for $T_C < 10^9$~K,
a neutron star would go through the pairing phase transition at ages
ranging from hundreds to tens of thousands of years, 
accompanied by a short phase of rapid cooling.  
This phenomenon,  illustrated in Fig.~6 of \cite{Page:2009ly}, 
closely resembles earlier results,
e.g., Fig.~1 of \cite{Gusakov:2004ys} and Fig.~8 of \cite{Yakovlev:2004kx}. 
In Fig.~\ref{Fig1}, we show similar results of models with 
the likely surface chemical composition of Cas A from
observations \cite{Ho:2009fk,Heinke:2010xy}.
The dotted curves show that rapid cooling is strongly dependent on
$T_C$, a value $\simeq 0.5\times 10^9$~K being favored by the
present age of Cas A.

%++++++++++++++++++++++++++++++++++++++++++++++++++++++++++++++++++++++
\begin{figure}[t]
   \begin{center}
   \includegraphics[width=0.48\textwidth]{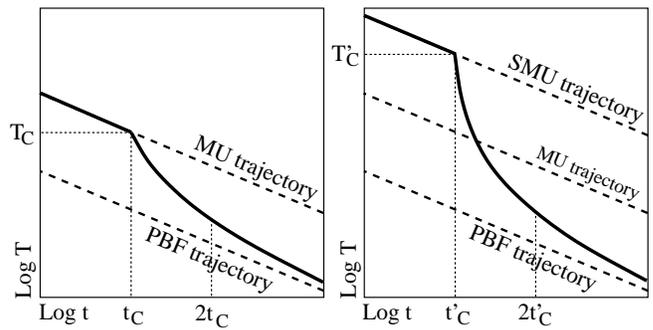}
  \end{center}  
   \caption{Schematic evolution of the internal temperature $T$ of a
                 neutron star (solid curves).  Left panel: Initially,
                 $T$ follows the upper modified Urca (MU) trajectory,
                 Eq. (\ref{Eq:T_MU}), until 
                 $T$ reaches  $T_C$ at time
                 $t_C$ when the pair breaking and formation (PBF) process turns on.
                 Thereafter, $T$ rapidly transits, on an initial time scale
                 $\tau_{_\mathrm{TR}} = t_C/f$, Eq. (\ref{Eq:T_PBF}),
                 toward the lower PBF trajectory, Eq. (\ref{Eq:T_PBF2}).  
                 Empirically, the rapid transition lasts up to a time $\lesssim 2t_C$.
                 Right panel: 
                 Evolution using a superconductivity
                 suppressed modified Urca (SMU) rate.
                 For the transit to start at a time $t'_C \approx t_C$
                 the trajectory requires a $T'_C$ larger than $T_C$ of the left panel.
                 The early transit has a shorter time scale
                 $\tau_{_\mathrm{TR}}' = t'_C/f'$ with $f' > f$,
                 but the late time evolution is almost identical to that of the left panel. 
                 The left (right) panel corresponds to models of Fig.~\ref{Fig1}
                 (Figs. \ref{Fig3} and \ref{FigMass}).  
                 }
   \label{Fig2} 
\end{figure}
%++++++++++++++++++++++++++++++++++++++++++++++++++++++++++++++++++++++

We now offer physical insight into the evolution of Cas~A
based on an analytical model developed in \cite{Page:2006ly}.
Assuming the star's core to be isothermal at a temperature $T$, 
a state reached within a few years after birth, and that the dominant
cooling occurs through $\nu$-emission with a luminosity $L_\nu$, 
which is correct for up to several tens of thousands of years, energy
balance gives
\be
C_V \frac{dT}{dt} = - L_\nu \,,
\label{Eq:Cool}
\ee
where $C_V$ is the star's total specific heat.
Neglecting the effects of pairing \cite{Note1}, one can write
$C_V = C_9 \, T_9$ with
$C_9 \sim 10^{39}$~erg~K$^{-1}$ and 
$T_9 \equiv T/(10^9$~K).
Similarly, $L_\nu$ from the MU process can be written
$L_\mathrm{MU} = L_9 \, T_9^8$ with
$L_9 \sim 10^{40}$~erg~s$^{-1}$.
Once triggered, neutrino emission from the PBF process is more efficient than the MU process; 
i.e., $L_\mathrm{PBF} = f  L_\mathrm{MU}$
with $f \sim 10$ \cite{Note2}.
Values of $C_9$ and $L_9$ can be found in \cite{Page:2006ly} or 
from  Figs.~11 and 20 of \cite{Page:2004zr}.
Considering ages large enough such that $T$ is much smaller than the initial
temperature,  Eq. (\ref{Eq:Cool}) gives
\be
T  = 10^9 \, \mathrm{K} \left(\tau_{_\mathrm{MU}}/t\right)^{1/6}
\label{Eq:T_MU}
\ee
with a MU cooling time-scale
$\tau_{_\mathrm{MU}}~\equiv~10^9~C_9/6L_9~\sim~1 \mathrm{yr}$.
The temperature $T$ will reach $T_C$ at a time
 $t_C = \tau_{_\mathrm{MU}} /  T_{C \, 9}^6$.
If this happened recently in the core of Cas~A, i.e.,  at $t_C \simeq (0.5 - 0.9) \times 330$ yrs,
one deduces that
\be
T_{C} = \max T_{cn}(\rho) =
10^9 \, \mathrm{K} (\tau_\mathrm{MU}/t_C)^\frac{1}{6} \sim  0.5 \times 10^9 \, \mathrm{K}.
\label{Eq:Tc}
\ee
After that moment, $L_\nu$ rapidly increases, by a factor $f$, and the solution of Eq. (\ref{Eq:Cool}) switches to
\be
T = \frac{ T_C}{\left[1 + f(t-t_C)/t_C\right]^{1/6}} =
\frac{ T_C}{\left[1 + (t-t_C)/\tau_{_\mathrm{TR}}\right]^{1/6}}
\label{Eq:T_PBF}
\ee
with a transit time scale $\tau_{_\mathrm{TR}}$ given in terms of the PBF time scale 
$\tau_{_\mathrm{PBF}} = \tau_{_\mathrm{MU}}/f$ as
\be
\tau_{_\mathrm{TR}} \equiv  \frac{t_C}{f} =
\frac{\tau_{_\mathrm{PBF}}}{T_{C \, 9}^6} 
= \frac{\tau_{_\mathrm{MU}}}{f \, T_{C \, 9}^6} \,. 
\label{Eq:tau_PBF}
\ee
With an $f$ of the order of $10$, one can naturally expect a transit time scale of a few decades in the case of Cas~A.
When $t \gg t_C$, the solution in Eq. (\ref{Eq:T_PBF}) has the same form as that in Eq. (\ref{Eq:T_MU}), 
but with the shorter time scale $\tau_{_\mathrm{PBF}}$:
\be
T = 10^9 \, \mathrm{K}  \left( \tau_{_\mathrm{PBF}} / t \right)^{1/6} =
       10^9 \, \mathrm{K}  \left( \tau_{_\mathrm{MU}} / f t \right)^{1/6} \,.
\label{Eq:T_PBF2}
\ee
The above evolution 
is depicted schematically in the left panel of Fig.~\ref{Fig2}. 

\begin{figure}[t]
   \begin{center}
   \includegraphics[width=0.48\textwidth]{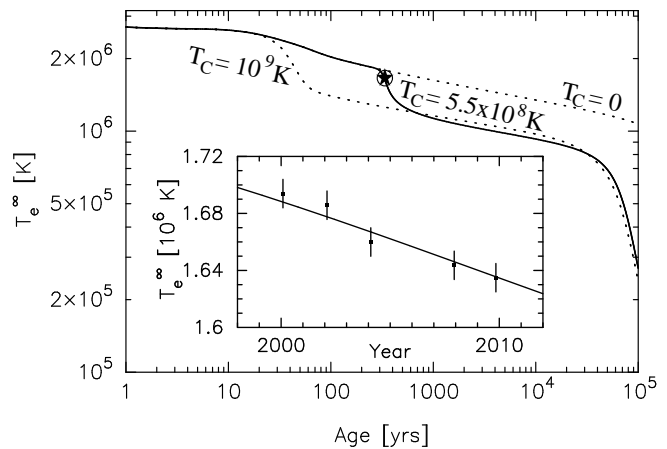}
  \end{center}
   \caption{A typical good fit to Cas A's rapid cooling for a 
                $1.4 {\rm M}_\odot$ star, built from the EOS of APR \cite{Akmal:1998qf} with an envelope
                mass $\Delta M_\mathrm{light} =  5\times 10^{-13} {\rm M}_\odot$.
                The two dotted curves, with indicated values of $T_C$, are to guide the eye.
                The three models have a proton $^1S_0$ gap from  \cite{Chen:1993ys}
                (the model ``CCDK" in \cite{Page:2004zr}) which results in 
                the entire core being superconducting.
                The insert shows a comparison of our results with the five data points of \cite{Heinke:2010xy}
                along with their $1\sigma$ errors.
                }
   \label{Fig3} 
\end{figure}

\begin{figure}[t]
   \begin{center}
   \includegraphics[width=0.40\textwidth]{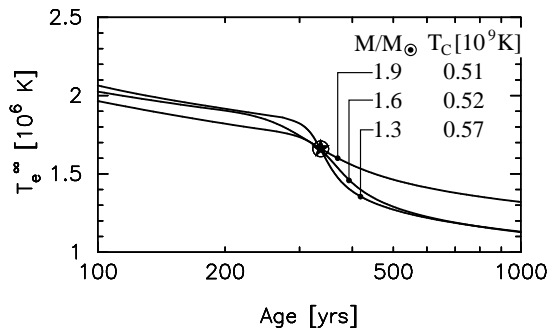}
  \end{center}
   \caption{Cooling curves with different masses and values of $T_C$ 
                  as indicated. 
                 	For the $1.9 {\rm M}_\odot$ star, 
                  $\Delta M_\mathrm{light} =  5 \times 10^{-11} {\rm M}_\odot$.
                  	For the other two masses shown, 
	                  $\Delta M_\mathrm{light} = 5\times 10^{-13} {\rm M}_\odot$.
	         The assumed proton  $^1S_0$ gap is the same as in Fig.~\ref{Fig3}.
	         The slopes, at the current age of Cas~A, are $-1.4$, $-0.9$, and $-0.5$
	         for the $1.3$, $1.6$, and $1.9 M_\odot$ models, respectively:
	         the decrease, with increasing mass, directly reflects the decrease of the
	         core's fractional volumes in which protons are superconducting.
                  }
   \label{FigMass} 
\end{figure}

For comparison with observations, we convert the internal $T$ 
to the observed effective $T_e$ using 
(see, e.g., \cite{Page:2006ly})
\be
T_e \simeq T_{e0} \, (T/10^8 \mathrm{K})^\beta \; \mathrm{K} \, ,
\label{Eq:TeTb}
\ee
where $T_{e0} \sim 10^6$ K and $\beta \sim 0.5$.  The evolution of
$T_e$ is hence similar to that of $T$, and the internal cooling curves
of Fig.~\ref{Fig2} map onto analogous models of Figs.~\ref{Fig1}, \ref{Fig3}, and \ref{FigMass}.  
The scale $T_{e0}$ and the exponent $\beta$ in
Eq.~(\ref{Eq:TeTb}) both depend on the chemical composition of the envelope. 
The presence of light elements, e.g., H, He, C, and/or O, 
increases $T_{e0}$ and reduces $\beta$
compared to the case of heavy elements, e.g., Fe,
depending on the total mass $\Delta M_\mathrm{light}$ of light elements \cite{Potekhin:1997kx}.

Using Eq.~(\ref{Eq:TeTb}), the slope
$s=d\log_{10}T_e/d\log_{10}t$ of the transit cooling curve from
Eq.~(\ref{Eq:T_PBF}) is
\be
s = \beta \frac{d\log_{10} T}{d \log_{10} t}
= -\frac{\beta}{6} \; \frac{f \, t/t_C}{1+f (t-t_C)/t_C} \,,
\label{Eq:slope}
\ee
whereas the slopes of 
the asymptotic trajectories,
Eqs.~(\ref{Eq:T_MU}) and (\ref{Eq:T_PBF2}), are both 
$s = -\beta/6 \sim -1/12$. 
As long as $t$ is only slightly larger than $t_C$, the transit
slope is larger than those of the asymptotic trajectories by a factor $ \sim f$.
The observed slope over a 10 yr interval is 
$s_\mathrm{obs} \simeq - 1.4$. 
Note, however, that the model "0.5" of Fig.~\ref{Fig1} does not exhibit such a large slope.
We are thus led to investigate the origin of the rapidity of Cas A's cooling.  

Several factors influence the rapidity of the transit phase.
Firstly, $L_\mathrm{PBF}$ depends on the shape of the
$T_{cn}(\rho)$ curve.  A weak $\rho$ dependence, i.e., a wide
$T_{cn}(\rho)$ curve, results in a thicker PBF neutrino emitting shell
and a larger $L_\mathrm{PBF}$ than a strong $\rho$ dependence.
Secondly, the $T$ dependence of $T_e$, i.e., the parameter $\beta$ in
Eq.~(\ref{Eq:TeTb}), also affects the slope in Eq.~(\ref{Eq:slope}).
Thirdly, protons in the core will likely exhibit superconductivity 
in the $^1S_0$ channel.
Most calculations of the proton critical
temperature, $T_{cp}(\rho)$, are larger than $T_{cn}(\rho)$ at low
densities.
Proton superconductivity suppresses
the MU process in a large volume of
the core at a very early age, reducing
$L_\mathrm{MU}$ \cite{Note3}.  
In our analytical model, this reduction translates to
a lower $L_9$ and, hence, to a larger $f$. 
The analytical model as well as
our calculations reveal that proton superconductivity significantly accelerates
cooling during transit and results in a large slope.
This feature, essential to account for Cas A's cooling rate,  
is illustrated in the right panel of Fig.~\ref{Fig2}.

By varying the relevant physical ingredients, such as the density range of proton
$^1S_0$ superconductivity, the shape of the $T_{cn}(\rho)$ curve, the
chemical composition of the envelope, and the star's mass, many models
can reproduce the average observed $T_e$ of Cas A.  These models yield
slopes ranging from $\sim -0.1$ (no rapid cooling and no constraint on
$T_C$) up to $-2$.  
A typical good fit to the rapid cooling of Cas~A is shown in Fig.~\ref{Fig3},
where the large slope results from the strong suppression of $L_\mathrm{MU}$
by extensive proton superconductivity.
Fig.~\ref{FigMass} demonstrates that the result $T_C\simeq 0.5\times 10^9$~K
does not depend on the star's mass, but that the slope during the transit is very
sensitive to the extent of proton superconductivity.
Models successful in reproducing the observed slope require
superconducting protons in the entire core.
Although spectral fits \cite{Ho:2009fk} seem to indicate that Cas A has a larger 
than canonical mass (1.4 M$_\odot$), a 
recent analysis  \cite{Yakovlev:2010fk}
indicates compatibility, to within $3\sigma$, with a smaller mass, 1.25 M$_\odot$.  
The need for extensive proton superconductivity to reproduce the 
large observed slope favors moderate masses
unless superconductivity extends to much higher densities than current models predict
(see, e.g., Fig 9 in \cite{Page:2004zr} for a large sample of current models).

%++++++++++++++++++++++++++++++++++++++++++++++++++++++++++++++++++++++
%++++++++++++++++++++++++++++++++++++++++++++++++++++++++++++++++++++++

The inferred  $T_C \simeq 0.5 \times 10^9$~K, either from
Figs.~\ref{Fig1}, \ref{Fig3}, and \ref{FigMass} or from
Eq.~(\ref{Eq:Tc}), appears quite robust and stems from
the small exponent in the relation
$T_C~\propto~(C_9 L_9^{-1}t_C^{-1})^{1/6}$.  
Assuming $L_9$ is not very strongly affected by proton superconductivity,
$T_C$ is mostly determined by $t_C$ which,
given the briefness of the rapid cooling transit phase,
cannot be much smaller than the age of Cas~A.
On the other hand, the rapidity of the cooling, i.e., 
the slope $s$, is predominantly controlled by proton superconductivity
reflected in the suppression of $L_\mathrm{MU}$.
We note, however, that if the proton $^1$S$_0$ gap 
extends to much higher densities than current theoretical models
indicate, suppression of $L_\mathrm{MU}$ could be larger than considered here.
This suppression can be up to a factor $\sim 50$; thus, 
 $T_C$ could almost reach $10^9$ K \cite{Note4}.

%++++++++++++++++++++++++++++++++++++++++++++++++++++++++++++++++++++++
%++++++++++++++++++++++++++++++++++++++++++++++++++++++++++++++++++++++

Although we have assumed the minimal cooling paradigm \cite{Page:2004zr},
our results remain valid in case some fast $\nu$ process is allowed.
Any of these processes is so efficient that the high observed $T_e^\infty$ of Cas~A
implies that it has been strongly suppressed at a very young age 
(likely by pairing of one of the participating particles) and rendered inoperative  \cite{Page:2006ly}.  
In such a case, $L_\nu$, prior to the onset of the neutron $^3$P$_2$ phase
transition, was dominated by the MU and nucleon bremsstrahlung
processes, exactly as in the minimal cooling paradigm, and the results
of our present analysis are not altered.

Our value of $T_C$ deduced from Cas A's cooling 
is compatible with the requirement $T_C\gtrsim 0.5 \times 10^9$ K 
inferred from the minimal cooling paradigm described in \cite{Page:2009ly}.
We predict that Cas~A will continue to cool for 
several decades with an almost constant rate, 
in contrast to two other interpretations \cite{Heinke:2010xy}; 
a relaxation from a recent sudden energy release
deep in the crust or a low level accretion from a fossil disc
that imply that the observed cooling is
episodic and that the light curve should flatten out soon. 
Distinguishing among these possibilities 
and our model will be possible by 
continuous monitoring of the thermal evolution of Cas A.

%%%%%%%%%%%%%%%%%%%%%%%%%%%%%%%%%%%%%%%%%%%%%%%%%%%%%%%%%%%

\begin{acknowledgments}
D.P.'s work is supported by a grant from UNAM-DGAPA \# IN122609.
MP and JML acknowledge research support from  
the U.S. DOE grants DE-FG02-93ER-40756 and DE-AC02-87ER40317, respectively.  
AWS is supported by the Joint Institute for Nuclear Astrophysics under
NSF-PFC grant PHYS 08-22648, by NASA ATFP grant NNX08AG76G, and by the
NSF under grant number PHY-08-00026.
\end{acknowledgments}

%%%%%%%%%%%%%%%%%%%%%%%%%%%%%%%%%%%%%%%%%%%%%%%%%%%%%%%%%%%

%%%%%%%%%%%%%%%%%%%%%%%%%%%%%%%%%%%%%%%%%%%%%%%%%%%%%%%%%%%

%%%%%%%%%%%%%%%%%%%%%%%%%%%%%%%%%%%%%%%%%%%%%%%%%%%%%%%%%%%
%%%%%%%%%%%%%%%%%%%%%%%%%%%%%%%%%%%%%%%%%%%%%%%%%%%%%%%%%%%
\end{document}